\documentclass[letterpaper, 10 pt, conference]{ieeeconf}  

\IEEEoverridecommandlockouts                              
\usepackage{float}
\usepackage{multirow}
\usepackage{rotating}
\usepackage{booktabs}
\usepackage{amsmath,amsfonts,amssymb}
\usepackage{url}
\usepackage{slashbox}
\usepackage{math}
\usepackage{hyperref}

\usepackage{xcolor}
\usepackage{color, colortbl}

\newcommand{\tr}{\textrm{t}}

\definecolor{gr}{gray}{0.70}

\title{\LARGE \bf
Sound Representation and Classification Benchmark \\
for Domestic Robots
}

\author{Janvier Maxime$^{\dag}$, Xavier Alameda-Pineda$^{\dag}$, Laurent Girin$^{\dag,\ddag,\#}$ and Radu Horaud$^{\dag}$ \\
	$^{\dag}$INRIA Grenoble Rhone-Alpes, $^{\ddag}$GIPSA-LAB and $^{\#}$Universit{\'e} Grenoble Alpes
\thanks{This work is financially supported by the ``Direction G{\'e}n{\'e}rale de l'Armement'' (DGA), The French Government Defense}
}

\begin{document}

\maketitle
\thispagestyle{empty}
\pagestyle{empty}

\begin{abstract}
We address the problem of sound representation and classification and present results of a comparative study in the context of a
domestic robotic scenario. A dataset of sounds was recorded in realistic conditions (background noise, presence of several sound sources, reverberations,
etc.) using the humanoid robot NAO. An extended benchmark is carried out to test a variety of
representations combined with several classifiers. We provide results obtained with the annotated dataset and we assess the methods quantitatively on the basis of their classification scores, computation times and memory requirements. The annotated dataset is publicly available at https://team.inria.fr/perception/nard/.
\end{abstract}

\section{Introduction}

In order to naturally interact with objects and people, robots need robust and efficient perception
capabilities. For example, human-robot interaction requires the recognition of gestures, actions,
and facial expressions. There has been tremendous progress  towards endowing
robots with visual perception. Nevertheless, the visual modality has its own limitations, e.g., it
cannot operate in bad (too dark or too bright) lighting conditions, and the interaction is inherently limited to objects
and people that are within the visual field. In parallel to visual information, \emph{sounds} produced by objects, by humans, or human-object interactions convey rich cognitive information about the ongoing context, events, and communicative behaviors. 

Compared to visual analysis, audio analysis is complementary but it also has its own advantages. Visual data are huge, visual information is complex to extract, and hence efficient visual routines may be difficult to embed into the robot's onboard hardware/software resources. In contrast, acoustic signal processing may be quite efficient, because the lower amount of data to be analyzed (depending
however on the complexity of the acoustic scene). By using hearing, a robot may be able to
recognize the ongoing events, estimate their relevance, and take appropriate decisions, even if they are not within the range of the visual sensors. Moreover, proper recognition and localization of sound events may be used to trigger visual attention mechanisms.

\begin{figure}[t!]
 \centering
 \includegraphics[width=0.8\linewidth]{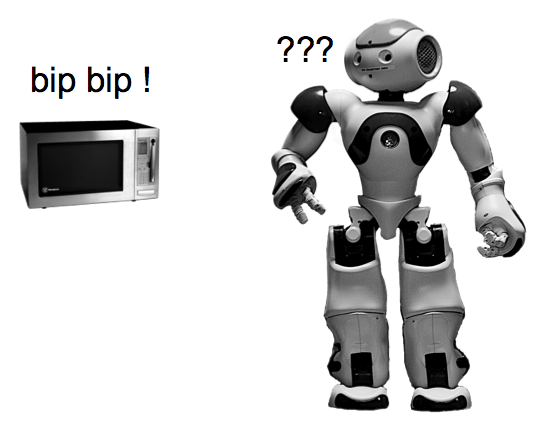}  
\caption{Domestic robots, such as NAO, should be able to robustly recognize sounds in the presence of room reverberations, background noise and competing sound sources.}
 \label{NAO_Bip}
\end{figure}

Therefore, audition is considered with increasing attention by robotic practitioners
since hearing capabilities are likely to considerably improve the overall ``cognitive understanding'' of a scene as an extended
catalogue of events, and improve the interactive capabilities of robots with humans, as well as with animals and objects, including
other robots. This is related to computational auditory scene analysis (CASA)  which attempts to model the abilities of  human audition, notably to
segregate coherent auditory streams~\cite{bregman1994}. It covers a set of challenging problems some of which have already been successfully investigated in robotics: multiple-source localization \cite{Huang1999} and separation
\cite{nakadai2002}, speech recognition \cite{yamamoto2006}, speech/non-speech/music classification,
detection/segmentation and recognition of elementary sounds (possibly in background signal/noise), etc. Some of these modules were successfully  integrated in robotic platforms, e.g., HARK \cite{nakadai2008open} and ASIMO
\cite{sakagami2002}, to cite just a few.

In the framework of robot audition, this paper addresses isolated recognition of ``domestic sounds''. We address both audio-signal representation and classification. The audio recordings are collected with 
a NAO robot manufactured by Aldebaran-Robotics\footnote{http://www.aldebaran-robotics.com}.  Similar benchmarks can be found for example in \cite{chu2006}
(for scene recognition), \cite{sasaki2009}, \cite{yamakawa2011} and \cite{stork2012}.
This setup implies notable difficulties, the most notable one being the low microphone quality currently available with NAO. The collected sounds are from a real-world scenario, e.g., fig.~\ref{NAO_Bip}: there are different types of sound sources, located at different (more or less distant) positions relatively to the robot head. The recorded audio signals are perturbed by room reverberations and by various linear or non-linear filtering effects (notably the robot's head-related transfer function which is difficult to estimate). The sounds are corrupted by the internal noise coming from the hardware inside the robot head. Also, the robot has limited computational capabilities, and this is expected to have a strong influence on the choice of signal representation and classification algorithms, as detailed below. 

The experimental data and used in this paper stays in contrast with clean sound databases recorded with high-quality microphones in specially equipped rooms. Moreover, automatic speech recognition (ASR) techniques often use close-range microphones which is not the case here since the robot is at some distance from the audio sources. 

We consider short sounds, typically in the range 0.1 to 1.0 seconds, that result from such events as the opening/closing of a
door, people dropping an object or clapping hands, as opposed to continuous sounds or continuous sound streams. Many of
these short sounds have an impulsive nature, and they are assumed to have well-defined start- and end-points. Therefore,
basic detection techniques based on signal energy or other statistics can be used to pre-segment the signal before
classification \cite{janvier2012}, and we do not address the detection/segmentation problem in the present paper: we
assume that a correct segmentation of these short sounds is available. We also assume that the sounds do not overlap
in time. Non-stationary sound streams such as continuous speech or music signals are not considered in the present study
(although our dataset contains isolated spoken words, see Section~II). Continuous speech is usually processed with specific classifiers, e.g., hidden
Markov models (HMMs), that model the dynamic evolution of the spectral patterns corresponding to the successive phonemes
\cite{rabiner1989}. Music signals are particularly tricky to process because of the richness of their content. More
stationary sound streams such as the flow of tap water, washing machine, fans, etc., are not considered as well. The
latter category can be considered either as long sound events or as background noise/context for overlapping short sound events. All
these problems will be considered in future extensions of the present work which focuses on implementation of short
sounds recognition in a robotic context. Note that this task is not trivial in itself, even without the limitations of
the robotic context, depending on the number and complexity of the sound categories. For example, different objects can
produce similar sounds that should, or should not, be classified together depending on the application. On the opposite,
the same physical object can produce different types of sounds that may not belong to the same category. Our dataset contains 42 sound categories, which  is a quite substantial number of sound types, as compared to previous
studies, e.g., 10 as in \cite{sasaki2009,tran2011,toyoda2004,yamakawa2011}, 15-16 as in \cite{guo2003,ramasubramanian2011} or 22
categories \cite{stork2012} .

Our main goal is to carry out a benchmark assessing different signal representations (audio features) and different classifiers, in the spirit of what was done in, e.g.,~\cite{cowling2003} for environmental sound recognition. We selected several feature spaces to represent sounds, as well as a number of classification techniques. Many possible combinations of features and classifiers were tested, possibly to reveal general trends and propose an optimal solution. 

Obviously, the accuracy score is the most important gauge for a classifier. The tested techniques are dedicated to be
embedded in autonomous robots, hence other important indicators are analyzed and reported. First, robots have to work in
(quasi) real-time, therefore execution has to be as fast as possible. Three time statistics are provided: the
\textit{feature computation time} (time to compute features from a raw signal of a given length),
\textit{training time} (time to train all the models for classification), and \textit{recognition time} (time to
classify a new incoming sound of a given length). Secondly, memory requirement is also a valuable resource in an embedded
system, and we estimate the \textit{training memory} (memory used to store the trained models). Getting the accuracy score, the
computation times and memory costs for each feature/classification method will allow us to find optimal solution(s) or
good trade-offs for reliable sound recognition with a consumer robot.

The remainder of this paper is organized as follows. Section~\ref{sec:data} describes in detail the dataset recorded and used in this study. Sections~\ref{sec:features} and~\ref{sec:classifiers} present respectively the different features and classifiers that were used. Experiments and results are presented in Section~\ref{sec:experiments}. Conclusions and the future work are expanded in Section \ref{sec:futurework}.

\section{The Data}
\label{sec:data}

The dataset must have the following characteristics: (i) to be recorded with low-quality sensors, (ii) to suffer from typical
internal robot noise, (iii) to be recorded in realistic domestic environments, i.e., in rooms with no special
acoustic characteristics, presence of reverberations and of multiple sound-source randomly distributed across the room, and
(iv) containing a substantial number of real-world sound types with only a few samples per class. Up to our knowledge, no
existing database that fulfills these requirements is available. Therefore, we recorded a database by placing NAO in both a
home and an office, and by using its frontal 300Hz -- 18kHz bandpass microphone. The collected signals are sampled at 48kHz and quantized
at 16 bits per sample. The robot-head fan produces noise within the band from 0 to 4~kHz, shading weak sounds. During
recording, the robot stands still and hence is not affected by noise generated by its motion. The dataset is available
online \footnote{https://team.inria.fr/perception/nard/}. Four scenarios and 42 sound classes were
considered, as summarized in Table~\ref{tab:data_sets}.

\begin{itemize}
\item \textbf{Kitchen}: The first part contains a large variety of everyday sounds collected in a home kitchen. We recorded 12 sound categories with different temporal and spectral characteristics: impulsive sounds (\textit{Close the microwave}, \textit{Choking}), harmonic sounds (\textit{Microwave alarm}) and transient sounds (\textit{Running the tap}, \textit{Eating}). The sounds were recorded from three different positions, 1 to 5 meters range and at various angles from the sound source. At each position, seven instances of each class were recorded, which sums up to 21 examples per class.

\item \textbf{Office}: The second part is related to an office environment. We acquired seven sounds: \textit{Door close, Door open, Door key, Door knock, Ripped Paper, Zip, (another) Zip.} They were randomly recorded from 0.3 to 5 meters range and from various angles. All the sound related to door actions were recorded using different doors. 
  
\item \textbf{Non-verbal}: The third part of the data contains non-verbal sounds, which are produced by humans, and can be seen as communication signals, but typically not taken into account in ASR systems. There are three classes (\textit{Fingerclap, Handclap, Tongue clic}) recorded from 0.3 to 5 meters range and from various angles, with four different people.

\item \textbf{Speech}: The fourth part of the dataset contains occurrences of isolated words. Even if speech recognition
is not in the scope of the present work, we judged of great interest to test methods designed for short sounds
recognition on such speech samples. Hence, we recorded twenty word classes from four different people placed in front
of NAO, roughly one meter away.
\end{itemize}

\begin{figure}[thpb]
  \centering
  \framebox{\parbox{3in}{\includegraphics[width=\hsize]{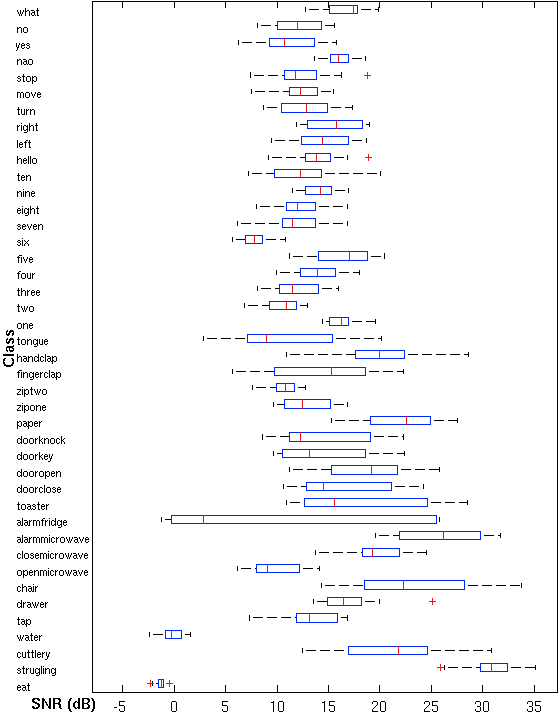}}}   
  \caption{Signal-to-Noise Ratio (SNR) per class. For each box, the central red mark denotes the median, the edges the 25th and 75th percentiles, the whiskers extend
to the most extreme data points not considered outliers, and outliers are plotted with a cross.}
  \label{SNR_dataset}
\end{figure}

Except for the \emph{Kitchen} classes, each class has 20 instances which made a total number of 852 sounds recorded for
the whole dataset. Considering that detection step is not addressed in this study, each sound has been manually
segmented using an audio editor. As an illustration of the signals ``quality'', Fig.~\ref{SNR_dataset} shows the 
signal-to-noise ratio (SNR) statistics for each class, the noise being here the internal noise, measured during absence
of any external sound. 

\begin{table}[h]
  \centering
  \caption{Taxonomy of the recorded data set classes.}
  \label{tab:data_sets}
  \begin{tabular}{ccc}
\toprule
      \textbf{Scenarios} & \textbf{Taxonomy} & \textbf{Classes} \\
\midrule
	\multirow{5}{*}{\textbf{Kitchen}} & ``Mouth'' sound & \textit{Eating}, \textit{Choking} \\
		& Cooking  & \textit{Cuttlery}, \textit{Fill a glass}, \textit{Running the tap} \\
		& \multirow{2}{*}{Moving} & \textit{Open/close a drawer}, \textit{Move a chair}  \\
		&  &  \textit{Open microwave},\textit{Close microwave} \\
		& Alarms & \textit{Microwave}, \textit{Fridge}, \textit{Toaster} \\
\midrule
	\multirow{2}{*}{\textbf{Office}} & Door & \textit{Close}, \textit{Open}, \textit{Key}, \textit{Knock} \\
	       & Others & \textit{Ripped Paper}, \textit{Zip}, \textit{(another) Zip} \\
\midrule
	\textbf{Nonverbal} &  & \textit{Fingerclap}, \textit{Handclap}, \textit{Tongue Clic} \\
\midrule
	\multirow{3}{*}{\textbf{Speech}} & Numbers & \textit{1,2,3,4,5,6,7,8,9,10} \\
		& \multirow{2}{*}{Orders} & \textit{Hello, Left, Right, Turn, Move} \\
		&  & \textit{Stop, Nao, Yes, No, What} \\
\bottomrule
  \end{tabular}
\end{table}

\vspace{-0.01cm}
\section{Audio Feature Representations}
\label{sec:features}
In this section, we present the different signal representations that were tested in our classification benchmark. Although quite short (see introduction), the considered signals are generally non-stationary, hence most of the features are actually \emph{time sequences} of \emph{feature vectors} computed using the very usual short-term sliding window approach widely used in audio processing. Except when specified, the window analysis is a 30ms Hamming window with 50\% overlap. All the features introduced in this section have been proposed in the audio processing literature \cite{peeters2004}.   

\subsection{Time-Domain Features}
\label{subsec:timefeatures}

\subsubsection{Energy}
We compute the energy as the root mean square of the samples in an audio frame (the rectangular window is used here). It can be seen a measure of amplitude variation over time.

\subsubsection{Zero Crossing Rate (ZCR)} defined as the number of zero crossings in an audio frame. It can be used to
classify voiced and unvoiced speech sounds, and it has also been used to differentiate speech, music and background
noise~\cite{saunders1996}.

\subsubsection{Sound Duration}
This feature is the total duration of the detected sound expressed in seconds. Therefore, in addition to being a scalar value, it is the only feature that is not extracted on a short-time basis. It may help to distinguish short, e.g. percussive, sounds from longer ones. 

\subsection{Frequency-Domain Features}
\label{subsec:freqfeatures}

All these features are computed using the Short-Term Fourier Tranform (STFT) of the signal.  $S(t,k)$ denotes the $k$-th \emph{magnitude} coefficient of the $N$-point STFT frame at time $t$.

\subsubsection{Spectral Roll-Off}
The Spectral Roll-off is the cut-off frequency below which 99\% the spectral energy is contained. It is used in speech recognition to classify voiced and unvoiced speech~\cite{scheirer1997}.

\subsubsection{Spectral Shape Statistics}
Those features characterize the overall shape of the spectrum using $n$-order moments of frequency bin weighted by spectral magnitude: 
$\mu_n(t) = \sum_{k=0}^{N-1} k^n S(t,k) \left/ \sum_{k=0}^{N-1} S(t,k) \right . $
The first moment, or spectral centroid or brightness, corresponds to the mean value of the weighted frequency. The second order moment measures the spread of the frequency distribution around the mean. The third order moment, or skewness, is a measure of the asymmetry of the distribution. The kurtosis (fourth order moment) is a measure of the ``peakedness'' of the distribution.


\subsubsection{Spectral Slope and Spectral Decrease}
The two features represents the global amount of decreasing of the spectral amplitude. The spectral slope is estimated by linear regression.
\[ S_{slope}(t) = \frac{ N\sum_{k}f_{t}(k)S(t,k) - \sum_{k}f_{t}(k)\sum_{k}S(t,k)}{N\sum_{k}f_{t}(k)^2-(\sum_{k}S(t,k))^2}, \]
where $f_k(t)$ represents the value of the linear regression at bin $k$ (and at time $t$).
The formulation of the spectral decrease comes from perceptual studies and tries to be coherent with human hearing~\cite{peeters2004}.
\[ S_{decrease}(t)= \frac{1}{\sum_{k=1}^{N-1}S(t,k)}\sum_{k=1}^{N-1}\frac{S(t,k)-S(t,0)}{k}. \]

\subsubsection{Spectral Flatness}
An estimation of the flatness of the magnitude spectrum is obtained by the ratio between its arithmetic and geometric mean (flat if $\approx1$ or peaky if $\approx0$): 
\[ S_{flat}(t) = \exp\left(\frac{1}{N}\sum_{k=0}^{N-1} \log(S(t,k))\right) \left/ \frac{1}{N}\sum_{k=0}^{N-1} S(t,k) \right.. \]

\subsubsection{Spectral Flux and Spectral Correlation}
The two features measure the average variation of spectral coefficients between two consecutive frames:
\[ S_{flux}(t) = \frac{\sum_{k}(S(t,k) - S(t-1,k))^2}{\sqrt{\sum_{k}S(t,k)^2}\sqrt{\sum_{k}S(t-1,k)^2}} \]
\[ S_{cor}(t) = \frac{\sum_{k}S(t,k)S(t-1,k)}{\sqrt{\sum_{k}S(t,k)^2}\sqrt{\sum_{k}S(t,-1k)^2}}.\]

\subsection{Mel-Frequency Cepstral Coefficients}
Widely used in speech and speaker recognition \cite{rabiner1989}, MFCCs are cepstral coefficients that represent the spectrum envelope on a perceptive mel-frequency scale. Those coefficients are computed as the discrete cosine transform (DCT) of the logarithm of FFT power coefficients passed through a mel-filter bank (40 log-spaced bands in the range 300Hz-10000Hz according to the following mel-scale  $1127\log(1 + f/700)$). Usually, the first coefficient is omitted and the first and second derivatives of the remaining coefficients can be added.

\subsection{Wavelet Features}
The wavelet transform \cite{mallat1999} transpose a signal from time domain to time-frequency domain like the STFT
although the different family of basis functions, allowing multi-resolution analysis to get a variable time and
frequency resolution. The discrete version of the transform \cite{lin2005} uses a $M$-stage cascade of a downsampling by
2 and a high-pass and low-pass filter. Thus, a signal $x(n)$ can be decomposed on $a_i(k)$ and $d_i(k)$ with $i =
1,...,M$ called respectively the approximation coefficients and the details coefficients. Inspired from \cite{lin2005}
and \cite{tzanetakis2001}, the feature vector is the concatenation of the mean and the standard deviation of the
coefficients $a_M$ and $d_i$ with $i = 1,...,M$. The experiments use an 8th order decomposition on a 8-coefficient
Daubechies family. 

\subsection{Stabilized Auditory Images}
Based on modelling of the human cochlea, the auditory image model (AIM) of \cite{walter2011} produces stabilized
auditory images (SAI), which are a time delay-frequency sound representation close to a correlogram. The process chains
three main stages, multi-channel gammatone filter bank, half-wave rectification and triggered time integration, and
leads to a representation with high dimensionality. A technique was proposed in \cite{lyon2010} to reduce the
dimensionality of the SAI features. This procedure consists of three steps: create patches from the SAI, compute a
low-dimensional vector representation of each patch, and concatenate these patch feature vectors to form the final
feature vector.

\subsection{Post-processing}
Depending on the feature nature, the successive feature vectors $\xvect^t$ of a given sound can be further processed to produce different final features, which will feed the classifiers:
\begin{itemize}
\item The \emph{sequencing} i.e. simple concatenation, of the (original) successive vectors $\xvect=[\xvect^1,\ldots,\xvect^T]$.
\item The \emph{mean} of the vectors over the entire acoustic event. The concatenation of the mean and standard deviation can also be used.
\item The \emph{bag-of-words} (BoW) approach. The features of all sounds are first clustered using the $K$-means algorithm. Then, each sound has its feature vectors quantized using the resulting centroids, and is then represented as the \emph{normalized histogram} of centroid occurrences.
\item The \emph{interpolation} of the feature vector sequence to the mean duration $\overline{T}$ of all vector sequences in the database. Each sound is thus represented by $\overline{T}$ interpolated feature vectors sequenced into $\xvect_I=[\xvect_I^1,\ldots,\xvect_I^{\overline{T}}]$. 
\end{itemize}

The interpolation enables to normalize the vector sequence along the time axis, so that the new representation can be used by ``fixed data length'' classifiers. It amounts to a simplified Dynamic Time Warping (DTW) applied ``blindly'', i.e. without inspecting the fine structural organization of the sounds. The bag-of-words also intrinsically enables (temporal) normalization but without taking into account the timeline ordering of the vector sequence.

 Finally, it can also be noted that the final feature representation may also consist of the (row-wise) concatenation of
several different features. This is a particular (straightforward) case of information fusion for classification, a vast
domain which deepened investigation in the context of sound recognition by a robot is out of the scope of the present
paper.

\subsection{Implementations}
The computation of the wavelets has been done using the Matlab Wavelet Toolbox. SAI features are available at \cite{AIMC}. All other features have been computed with the Python/C++ toolbox YAAFE \cite{yaafe2010}. 

\section{Isolated Sound Classification}
\label{sec:classifiers}

In this section all the tested classifiers are described. A multiclass classifier consists of a mapping
$g:{\cal X}\times{\cal C}\rightarrow \mathbb{R}$, where ${\cal X}$ is the feature space, ${\cal C} = \{1,\ldots,C\}$ is
the set of labels and $C$ is the number of classes. The dimension of ${\cal X}$ may be fixed or varying with the sound,
depending on the feature used. Given a feature vector (or sequence of feature vectors) $\xvect\in{\cal X}$, $g(\xvect;c)$ is the score of classifying
$\xvect$ as $c$. The higher the score is, the more likely $c$ is the class of $\xvect$. Hence, a new unlabelled
observation $\xvect\in{\cal X}$ is classified as:
\[ c^*(\xvect) = \arg\max_{c\in{\cal C}} g(\xvect;c). \]
In the following, $\Xmat$ will denote the training set, i.e. a set of feature vectors
$\Xmat=\{\xvect^n\}_{n=1}^N$ which class is known, and that is used to train the classifiers.

\subsection{$K$-Nearest Neighbors}
The $k$-nearest neighbors ($k$-NN) classifier is based on the well-known $k$-NN algorithm which returns the subset of $S_k(\xvect)\subset\Xmat$, containing the $k$ closest points to a given vector $\xvect$. The mapping of the $k$-NN classifier is:
$g_{k\textrm{NN}}(\xvect,c) = \left|\{\tilde{\xvect}\in S_k(\xvect) | \mathfrak{c}(\tilde{\xvect}) = c\}\right|,$
where $\mathfrak{c}(\tilde{\xvect})$ means the class of $\tilde{\xvect}$. $g_{k\textrm{NN}}(\xvect,c)$ is the number of feature vectors among the $k$-nearest neighbors of $\xvect$ that belong to the class $c$. In other words, each of the $k$ neighbors votes for its own class, and the class with more votes is
assigned to $\xvect$.

\subsection{Quantized Nearest Neighbor}
The previous method needs to keep in memory all the training data during the recognition stage. QNN is able to
circumvent this issue by quantizing the features previously to the nearest neighbor search. More precisely, the vectors
are first divided in $P$ parts, leading to $P$ feature subspaces. If $\xvect_{n,p}$ denotes the $p$-th part of the
$n$-th training vector, we define $\Xmat_{,p} = \{\xvect_{n,p}\}_{n=1}^N$, the training set of the $p$-th feature
subspace. A $K$-means algorithm \cite{bishop2006} is ran for every $\Xmat_{,p}$, providing for a set of centroids. The quantization
function, that assigns the $p$-th subvector $\xvect_{,p}$ of $\xvect$ to its closest centroid is denoted by
$Q_p(\xvect_{,p})$. In that case the mapping $g$ is:
\[g_{\textrm{QNN}}(\xvect;c) = -\min_{\tilde{\xvect}\in\Xmat_c}
\left(\sum_{p=1}^P\|Q_p(\tilde{\xvect}_{,p})-Q_p(\xvect_{,p})\|\right)^\frac{1}{2},\]
where $\Xmat_c=\{\xvect\in\Xmat|\mathfrak{c}(\xvect)=c\}$. This corresponds to finding the quantized vector in the
training set closest to the quantized test vector, and assigning its class to $\xvect$. See \cite{janvier2012} for more
details on this technique. The method is parametrized by $K$ and $P$. The higher $K$ and $P$ are, the more costly the method is, and the higher the recognition rate is. Increasing $P$ may allow us to reduce $K$ with no negative effects on the recognition rate.

\subsection{Support Vector Machines}
The Support Vector Machines (SVM) is a discriminative binary classification method \cite{bishop2006}. It has been used in sound  recognition in multiple situations as in \cite{guo2003} and \cite{temko2006} with hierarchical structures or in \cite{asma2008} with 1-class SVMs. SVMs provides a discriminative function $h(\xvect)$, learnt form a set of positive examples and a set of negative examples. The points satisfying $h(\xvect)=0$ form a hyperplane in the space induced by the kernel function $k(\cdot,\cdot)$. $h(\xvect)>0$ means that $\xvect$ should be classified as positive and $h(\xvect)<0$ as negative. We refer the reader to \cite{bishop2006} for details on the formulation. Importantly, a
parameter ${\cal Q}$ regulates the amount of allowed misclassification in the training set, such that SVMs deal with
overlapping classes.

Since SVMs are binary classifiers, two strategies have been developed to use them in the multiclass task. On one hand
the \textit{one-versus-rest} (1vR), in which $C$ different SVMs are trained, one per class. In that case the mapping $g$
is defined as $g_{\textrm{1vR}}(\xvect;c)=h_c(\xvect)$ where $h_c(\xvect)$ is the discriminant function trained with $\Xmat_c$ and
$\Xmat\setminus\Xmat_c$. On the other hand the \textit{one-versus-one} (1v1) strategy, which corresponds to evaluate all
possible binary classification problems with $C$ classes. The classification mapping is then:
$g_{\textrm{1v1}}(\xvect;c) = \left|\{d\in{\cal C}|d\neq c , h_{c,d}(\xvect)>0\}\right|,$
where $h_{c,d}$ is the discriminant function trained with $\Xmat_c$ and $\Xmat_d$. As for $k$-NNs, this is
equivalent to say that each SVM is voting for one class and $\xvect$ is classified to the class with more votes. In our experiments, the 1V1 approach always outperformed 1VR both in terms of accuracy and speed, and we only consider 1V1 in the following.

Five different kernels are tested, namely: linear $k_L(\xvect,\yvect)=\xvect^\tr\yvect$, polynomial $k_P(\xvect,\yvect)=(\gamma \xvect^\tr\yvect+c_0)^d$, radial
basis $k_R(\xvect,\yvect)=\exp(-\gamma\|\xvect-\yvect\|^2)$, sigmoid $k_S(\xvect,\yvect)=\tanh(\gamma \xvect^\tr\yvect+c_0)$, and $k_{\chi^2}(\xvect,\yvect) = 1 - 2\sum_{i=1}^M\frac{(x_i-y_i)^2}{x_i+y_i}$, $M$ being the dimension of the features. The parameters of the SVMs are the misclassification regulation parameter ${\cal Q}$, the multiclass strategy, the kernel used and, if any, the kernel parameters.

\subsection{Gaussian Mixture Models}
\label{subsec:GMM}
The Gaussian Mixture Model (GMM) is a probabilistic generative model widely used in classification tasks. In our case,
we use one GMM per sound class. Each GMM is a weighted sum of $M$ Gaussian components (in this model, each observation
is assumed to be generated by one of these components), which parameter set denoted by $\lambda_c$ is composed of $M$
weights, mean vectors and covariance matrices. We learn $C$ sets of parameters $\lambda_c$, $C$ being the number of
classes using the well-known Expectation-Maximization (EM) algorithm. The mapping $g$ corresponds to the likelihood of
the observed data given the model parameters. For a sequence of feature vectors $\xvect=[\xvect^1,\ldots,\xvect^T]$,
which are assumed to be independent, we have: $g_\textrm{GMM}(\xvect;c) = \textrm{p}(\xvect|\lambda_c) = \prod_{t=1}^T
\textrm{p}(\xvect^t|\lambda_c).$ This method is parametrized by the number $M$ of Gaussians in the mixture, the maximum
number of EM iterations and the shape of the covariance matrices (full or diagonal). We refer the reader to
\cite{bishop2006} for more details about GMM.

\subsection{Hidden Markov Models}
The Hidden Markov Models (HMM) also belong to the family of generative models \cite{bishop2006,rabiner1989}. In a HMM the observations depend on a hidden discrete random variable usually called state, taking values from 1 to $S$. The probability of the observations given the state value is called emission probability. The state is assumed to be Markovian, that is, the state at time $t$ only depends on the state at time $t-1$. In addition, the states are constrained to happen in order, i.e. state $s$ before the state $s+1$; this is usually known as \textit{left-to-right} HMM. The emission probability is usually Gaussian or GMM. As in the case of GMM, one model $\xi_c$ per class is learnt (through an EM algorithm). The model consists of the parameters of the emission probability and the parameters modeling the markovian dynamics. The function $g$ is also the likelihood of the observations given the model: $g_\textrm{HMM}(\xvect;c)=\textrm{p}(\xvect|\xi_c).$ The parameters of the HMM are the parameters of the emission probability, the number of states $S$. We refer the reader to \cite{rabiner1989}  for more details about HMM.

\subsection{Implementations}
The $k$-NN, GMM algorithms comes from the Matlab toolboxes. The QNN algorithm is our own Matlab code  inspired from \cite{janvier2012}. The HMMs are developed using the machine learning PMTK3 library \cite{murphy2012}. The SVMs are implemented using libSVM \cite{libSVM}.

\section{Experiments}
\label{sec:experiments}

\label{sec:setup}

Given the database described in Section~\ref{sec:data}, a large set of combinations of feature types, features post-processing, and classifiers have been tested (note that all combinations do not make sense, e.g., some features are not appropriate for time interpolation; we implemented only relevant combinations). In order to be able to statistically compare the different sound recognition methods, we perform $k$-fold cross-validation repeated on $n$ different runs. The results are averaged on these $n$ runs, with $k$ and $n$ being set to $10$.

Tables~\ref{results:accuracy} to \ref{results:trainingcost} gather the different statistics on the different combinations of features (rows) + post-processing, and classifiers (columns). \emph{GMM-1} stands for the GMM method (section \ref{subsec:GMM}) applied when $T=1$, while \emph{GMM-T} corresponds to $T>1$. It is important to note that GMM-T and HMM methods are fed with sounds represented by the original (variable-length) sequence of feature vectors, whereas all the other classifiers are fed with a single fixed-size vector representation issued from post-processing by either mean (rows 1--4), Bag-of-Words (rows 5 and 6), or fixed-sized interpolation (rows 7 and 8). The latter still represents a vector time-sequence but of fixed length, and hence can be reshaped in a single vector. \emph{TTFF} stands for Time and Time-Frequency Features (corresponding to the features of section \ref{subsec:timefeatures} and  \ref{subsec:freqfeatures}). Cells filled with gray correspond to irrelevant combinations. 

\subsection{Results}
\label{sec:results}

Note first that the best results using \emph{TTFF} or by concatenating TTFF+MFCC have been found using the features \emph{Energy, ZCR, Spectral Decrease, Spectral Flatness, Spectral Slope}.  Therefore these features have been used in the presented results. Adding the \emph{Roll-off} and the \emph{Spectral Moments} gives similar results. The \emph{Sound Duration} is not a reliable feature in the present context, since it lead to drop in scores. 

As for accuracy, the best results are obtained using SVM classifiers on interpolated MFCC+TTFF coefficients (97\%
accuracy), followed by \textit{k}-NN with interpolated MFCC (96.2\%). HMM on MFCC coefficients, which is a very usual
combination in the literature, provides a very good baseline at 92.6\% good accuracy. Therefore, a major result here is
that, for short pre-segmented domestic sound recognition, a quite simple technique such as \textit{k}-NN, that requires
no training, can perform better than ASR reference methods such as HMMs. The latter requires both training and much
longer decoding time (see Table~\ref{results:recognitiontime}) and may be more appropriate for long and complex sound
sequences such as speech signals. As could be predicted, the preservation of dynamic information is important for
accurate recognition: see the 96.2\% good accuracy for \textit{k}-NN with MFCC + interpolation vs. 87.4\% for
\textit{k}-NN with MFCC + mean; see also the difference between GMM-1 and GMM-T. This is confirmed by the poor results
obtained with the Bag-of-Words approach which has not proven being relevant in these experiments (remind that BoW
histograms cumulate information over frames but loose the temporal structure; also, the histogram codebook cannot be
large because the training time grows up exponentially with $K$: for the experiments, we used $K=50$). However, accurate
vector alignment using advanced DTW as used in HMMs do not seem as crucial as for ASR: here basic fixed-size
interpolation seems efficient enough for the task at hand. This rises many questions about the (temporal and spectral)
structure of domestic sounds, that go beyond the scope of the present study. Waiting for further investigations, the
fact that \textit{k}-NN with simple feature sequence interpolation outperforms HMMs (and GMM-T) can be partly explained
by the fact that \textit{k}-NNs use original data in the recognition task while HMMs (and GMM-T) use data models. In
addition \textit{k}-NN is a discriminative technique, whereas HMMs (and GMM-T) are generative models. A consequence is
that \textit{k}-NNs have a very large memory requirement to store the prototypes (see Table~\ref{results:trainingcost}),
which a major drawback for autonomous robotics. 

Obviously, SVM is an interesting alternative, modestly increasing the recognition time over \textit{k}-NN for a much smaller memory cost. And so is QNN which has a larger recognition time but an even smaller memory cost. Therefore, the choice between \textit{k}-NNs, SVM and QNN should depend on the specifications of the autonomous robot in terms of computation and memory resources. GMM-T with MFCC has good accuracy performance but the recognition time is quite high, making it less interesting than the above-mentioned methods. It remains unclear why SVMs perform significantly better with MFCC+TTFF+interpolation than with MFCC+interpolation, whereas the difference is not so pronounced for \textit{k}-NN and QNN (and for some other settings, adding TTFF even decreases the accuracy scores; this is difficult to explain, except appealing to the redundancy between some TTFF features and MFCC information). 
Anyway, a major point that arises from this study is that, for short domestic sounds recognition, the three methods \textit{k}-NNs, SVM and QNN, combined with simple time interpolation of features, seem preferable to the (more complex) HMMs widely used for speech recognition and recently extended to the more general problem of sound scene analysis. 

To complement those results, we present in Table \ref{results:featuretime}, the time to compute feature vector(s) from a
sound (mean or sequence; column \emph{Feature}). In the column \emph{BoW}, \emph{K-means} is the training time of the
codebook, and \emph{Histo} the time to transform the feature vector(s) of one sound into a histogram.
\emph{Interpolation} is the time to perform the fixed-length time interpolation on the feature vector(s) of one sound.
We can see that the time to compute the feature vector (sequence) is reasonable but not negligible: for example, it is
an order of magnitude larger than the recognition time of \textit{k}-NN, but it is also more than an order of magnitude
lower than the recognition time for HMMs. For MFCC coefficients, the time needed to interpolate the MFCC sequence is
comparable to the time needed to calculate the coefficients. Note that the memory cost for training the models from data
are not considered in the present study, since this can be processed offline.

\section{Conclusions and  Future Work}
\label{sec:futurework}

We addressed the problem of sound recognition by an autonomous humanoid robot, by benchmarking a large set of audio feature representations, post-processing, and classification techniques. A major result of this work is that, for the 42 classes of kitchen/office/voice sounds that we considered, very good accuracy scores (larger than 92\% and up to 97\%) were obtained for three techniques of very reasonable complexity (at least for decoding), namely \textit{k}-NN, SVM and QNN. Moreover these methods were applied successfully on fixed-size sequences of MFCC vectors obtained  with very simple DTW (fixed-size interpolation). The performance in accuracy is of the same order and even outperforms the performance of HMMs applied on the original vector sequences, whereas the decoding time (hence computational cost) is much lower. Therefore, these three methods seem to be appropriate within the context a robotic implementation. 

A more thorough analysis of the nature of domestic sounds must be carried out to reveal if they are characterized by an
inner structure, in a similar way as, e.g. speech signals are characterized by successive phonemes (and transitions
between them). Domestic/environmental sounds can also be analyzed in terms of taxonomy, nature (matter of the object that generated the 
sound: metal, wood, glass, etc.), interactions or dynamics (friction, shock, etc.). To reach this goal, the number of
classes must be increased radically to reach several hundreds. The introduction of a ``garbage class'' is absolutely necessary, since, it is impossible to consider all the possible sound categories. Future work will also consider the
processing of continuous audio streams, e.g., taking into account stationary and less stationary background noise, or
"longer" sounds indicating a specific activity (e.g., tap water flushing). In addition to external noise, we will
address the problem of ego-noise (generated by robot joints in motion) detection and removal, as in
\cite{ito2005internal}. In the long run, we aim at merging the sound recognition system in a complete framework for
acoustic scene analysis including source localization and separation, embedded in the robot NAO.

\begin{table}[h]
  \caption{Accuracy Rates (in \%).}
  \centering
  {\scriptsize\begin{tabular}{c c c c c c c}
    \toprule
      
		     & $k$NN & QNN  & GMM-1 	      & GMM-T 		& HMM 			& SVM  \\
    \midrule
	TTFF 	     & 65.9  & 62.8 & 67    	      & 71   		& \cellcolor{gr} 	& 74.3 \\
	MFCC 	     & 87.4  & 82.1 & 89.5  	      & 95.4 		& \textbf{92.8} 	& 91.5 \\
	MFCC+TTFF     & 88.4  & 77.7 & 76.4  	      & 88   		& 92.2 			& 91.3 \\
	Wavelets     & 60.5  & 58.4 & 63    	      & 36.4 		& 61.3 			& 57   \\
	MFCC+BoW     & 55.8  & 53.9 & 45.2  	      & \cellcolor{gr}  & \cellcolor{gr}   	& 52.6 \\
	MFCC+TTFF+BoW & 60    & 55.8 & 41.1  	      & \cellcolor{gr}  & \cellcolor{gr}   	& 62.5 \\
	MFCC+Interp  & \textbf{96.2}  & 95.7 & \cellcolor{gr}  & \cellcolor{gr}  & \cellcolor{gr}   	& 92.3 \\
    MFCC+TTFF+Interp  & 94.1  & 94.2 & \cellcolor{gr}  & \cellcolor{gr}  & \cellcolor{gr}   	& \textbf{97}   \\
	SAI 	     & 83    & 80   & \cellcolor{gr}  & \cellcolor{gr}  & \cellcolor{gr} 	& 87   \\
    \bottomrule
  \end{tabular}}
  \label{results:accuracy}
\end{table}

\begin{table}[h]
  \caption{Training Time (in s).}
  \centering
  {\scriptsize\begin{tabular}{c c c c c c c}
    \toprule
		     & $k$NN 	      & QNN  & GMM-1 		& GMM-T 		& HMM  			& SVM   \\
    \midrule
	 TTFF 	     & \cellcolor{gr} & 0.6  & 0.3   		& 10.8  		& \cellcolor{gr}  	& 0.076 \\
	 MFCC 	     & \cellcolor{gr} & 1.1  & 0.4   		& 9.3   		& 14.2 			& 0.150 \\
	 MFCC+TTFF    & \cellcolor{gr} & 0.6  & 0.350 		& 7.6   		& 24.8 			& 0.092 \\
	 Wavelets    & \cellcolor{gr} & 1.3  & 1     		& 7.6   		& 52   			& 0.065 \\
	MFCC+BoW     & \cellcolor{gr} & 1.5  & 0.360 		& \cellcolor{gr}   	& \cellcolor{gr}   	& 0.350 \\
	MFCC+TTFF+BoW & \cellcolor{gr} & 1.5  & 0.5     		& \cellcolor{gr}   	& \cellcolor{gr}   	& 0.380 \\
	MFCC+Interp  & \cellcolor{gr} & 7.7  & \cellcolor{gr} 	& \cellcolor{gr} 	& \cellcolor{gr}   	& 2     \\
    MFCC+TTFF+Interp  & \cellcolor{gr} & 8.4  & \cellcolor{gr}   & \cellcolor{gr}  	& \cellcolor{gr}   	& 2.3   \\
	 SAI 	     & \cellcolor{gr} & 27.4 & \cellcolor{gr} 	& \cellcolor{gr} 	& \cellcolor{gr}   	& 0.79  \\
    \bottomrule
  \end{tabular}}
  \label{results:trainingtime}
\end{table}
\begin{table}[h]
  \caption{Recognition Time (in ms).}
  \centering
  {\scriptsize\begin{tabular}{c c c c c c c}
    \toprule
		        & $k$NN & QNN & GMM-1 		& GMM-T 	 & HMM 			& SVM \\
    \midrule
	TTFF 	 	& 0.3 	& 1   & 18   		& 36.3 		 & \cellcolor{gr}  	& 0.1 \\
	MFCC 	 	& 0.3 	& 1   & 19.7 		& 45.8 		 & 89 			& 0.2 \\
	MFCC+TTFF 	& 0.3 	& 1   & 20   		& 31   		 & 92 			& 0.2 \\
	Wavelets  	& 0.1 	& 1   & 2.1  		& 3    		 & 10 			& 0.1 \\
	MFCC+BoW     	& 0.4  	& 1.4 & 19   		& \cellcolor{gr} & \cellcolor{gr} 	& 8.2 \\
	MFCC+TTFF+BoW 	& 0.3   & 1.3 & 19    		& \cellcolor{gr} & \cellcolor{gr} 	& 0.9 \\
	MFCC+Interp     & 1.5   & 9   & \cellcolor{gr} 	& \cellcolor{gr} & \cellcolor{gr} 	& 2.3 \\
    MFCC+TTFF+Interp     & 1.5	& 9.4 & \cellcolor{gr}  & \cellcolor{gr} & \cellcolor{gr}   	& 2.5 \\
	SAI 	 	& 1.2 	& 5.2 & \cellcolor{gr} 	& \cellcolor{gr} & \cellcolor{gr} 	& 2.2 \\
    \bottomrule
  \end{tabular}}
  \label{results:recognitiontime}
\end{table}

\begin{table}[h]
  \caption{Memory needed to store the trained classifiers (in kB).}
  \centering
  {\scriptsize\begin{tabular}{c c c c c c c}
    \toprule
		        & $k$NN& QNN 	& GMM-1 	 & GMM-T 	  & HMM 	   & SVM \\
    \midrule
	TFF  	     	& 370  & 6 	& 39 		 & 130    	  & \cellcolor{gr} & 520  \\
	MFCC 	     	& 430  & 6 	& 50  		 & 180   	  & 1100  	   & 540  \\
	MFCC+TFF     	& 460  & 4 	& 46 		 & 200  	  & 1300  	   & 680  \\
	Wavelets  	& 350  & 10  	& 58 		 & 65 	 	  & 430 	   & 330  \\
	MFCC+BoW     	& 38   & 11.2   & 10.6       	 & \cellcolor{gr} & \cellcolor{gr} & 271  \\
	MFCC+TFF+BoW 	& 31   & 16.5   & 52.1   	 & \cellcolor{gr} & \cellcolor{gr} & 206  \\
	MFCC+Interp     & 5300 & 715    & \cellcolor{gr} & \cellcolor{gr} & \cellcolor{gr} & 2100 \\
    MFCC+TFF+Interp     & 6100 & 967    & \cellcolor{gr} & \cellcolor{gr} & \cellcolor{gr} & 3800 \\
	SAI 	     	& 4230 & 593 	& \cellcolor{gr} & \cellcolor{gr} & \cellcolor{gr} & 5550 \\
    \bottomrule
  \end{tabular}}
  \label{results:trainingcost}
\end{table}

\begin{table}[h]
  \caption{Feature Computation Time (in ms).}
  \centering
  {\scriptsize\begin{tabular}{c c c c c c c}
    \toprule
      &  \multirow{2}{*}{Feature} 	& \multicolumn{2}{c}{BoW} 		& Interpolation  \\
      &  			  	& $K$-means 	 & Histo.	    	& 		 \\
    \midrule
	TTFF  	 & 3   			& \cellcolor{gr} & \cellcolor{gr}  	& \cellcolor{gr} \\
	MFCC 	 & 2.4 			& 12.3	 	 & 0.8 			& 2.3		 \\
	MFCC+TTFF & 5.4 			& 13.3		 & 0.9		 	& 2.7		 \\
	Wavelets & 9.6 			& \cellcolor{gr} & \cellcolor{gr} 	& \cellcolor{gr} \\
	SAI 	 & 350 			& \cellcolor{gr} & \cellcolor{gr} 	& \cellcolor{gr} \\
    \bottomrule
  \end{tabular}}

  \label{results:featuretime}
\end{table}

\bibliographystyle{IEEEtran}

\end{document}